\documentclass[aip,jcp]{revtex4-1}

\usepackage[version=3]{mhchem} 
\usepackage[export]{adjustbox}

\newif\ifoutline
\outlinefalse
\begin{document}

\title{L-Edge Spectroscopy of Dilute, Radiation-Sensitive Systems Using a Transition-Edge-Sensor Array}

\author{Charles J. Titus}
\email{ctitus@stanford.edu}
\affiliation{Department of Physics, Stanford University, Stanford, California 94305, USA}
\author{Michael L. Baker}
\affiliation{Department of Chemistry, Stanford University, Stanford, California 94305, USA}
\author{Sang Jun Lee}
\affiliation{Stanford Synchrotron Radiation Lightsource, SLAC National Accelerator Laboratory, Menlo Park, California 94025, USA}
\author{Hsiao-mei Cho}
\affiliation{SLAC National Accelerator Laboratory, Menlo Park, California 94025, USA}
\author{William B. Doriese}
\author{Joseph W. Fowler}
\affiliation{National Institute of Standards and Technology, Boulder, Colorado 80305, USA}
\author{Kelly Gaffney}
\affiliation{Stanford Synchrotron Radiation Lightsource, SLAC National Accelerator Laboratory, Menlo Park, California 94025, USA}
\author{Johnathon D. Gard}
\author{Gene C. Hilton}
\affiliation{National Institute of Standards and Technology, Boulder, Colorado 80305, USA}
\author{Chris Kenney}
\affiliation{SLAC National Accelerator Laboratory, Menlo Park, California 94025, USA}
\author{Jason Knight}
\affiliation{Stanford Synchrotron Radiation Lightsource, SLAC National Accelerator Laboratory, Menlo Park, California 94025, USA}
\author{Dale Li}
\affiliation{SLAC National Accelerator Laboratory, Menlo Park, California 94025, USA}
\author{Ron Marks}
\affiliation{Stanford Synchrotron Radiation Lightsource, SLAC National Accelerator Laboratory, Menlo Park, California 94025, USA}
\author{Michael P. Minitti}
\affiliation{Linac Coherent Light Source, SLAC National Accelerator Laboratory, Menlo Park, California 94025, USA}
\author{Kelsey M. Morgan}
\author{Galen C. O'Neil}
\author{Carl D. Reintsema}
\author{Daniel R. Schmidt}
\affiliation{National Institute of Standards and Technology, Boulder, Colorado 80305, USA}
\author{Dimosthenis Sokaras}
\affiliation{Stanford Synchrotron Radiation Lightsource, SLAC National Accelerator Laboratory, Menlo Park, California 94025, USA}
\author{Daniel S. Swetz}
\author{Joel N. Ullom}
\affiliation{National Institute of Standards and Technology, Boulder, Colorado 80305, USA}
\author{Tsu-Chien Weng}
\affiliation{Stanford Synchrotron Radiation Lightsource, SLAC National Accelerator Laboratory, Menlo Park, California 94025, USA}
\author{Christopher Williams}
\affiliation{Department of Physics, Stanford University, Stanford, California 94305, USA}
\author{Betty A. Young}
\affiliation{Department of Physics, Santa Clara University, Santa Clara, California 95053, USA}
\author{Kent D. Irwin}
\affiliation{Department of Physics, Stanford University, Stanford, California 94305, USA}
\affiliation{SLAC National Accelerator Laboratory, Menlo Park, California 94025, USA}
\author{Edward I. Solomon}
\affiliation{Department of Chemistry, Stanford University, Stanford, California 94305, USA}
\affiliation{Stanford Synchrotron Radiation Lightsource, SLAC National Accelerator Laboratory, Menlo Park, California 94025, USA}
\author{Dennis Nordlund}
\affiliation{Stanford Synchrotron Radiation Lightsource, SLAC National Accelerator Laboratory, Menlo Park, California 94025, USA}

\begin{abstract}
We present X-ray absorption spectroscopy and resonant inelastic X-ray scattering (RIXS) measurements on the iron L-edge of 0.5 mM aqueous ferricyanide. These measurements demonstrate the ability of high-throughput transition-edge-sensor (TES) spectrometers to access the rich soft X-ray (100--2000~eV) spectroscopy regime for dilute and radiation-sensitive samples. Our low-concentration data are in agreement with high-concentration measurements recorded by conventional grating-based spectrometers. These results show that soft-X-ray RIXS spectroscopy acquired by high-throughput TES spectrometers can be used to study the local electronic structure of dilute metal-centered complexes relevant to biology, chemistry, and catalysis. In particular, TES spectrometers have a unique ability to characterize frozen solutions of radiation- and temperature-sensitive samples.

\end{abstract}

\maketitle

\section{Introduction}

Transition metals play critical roles in industrial and biological processes. The function of many catalysts is determined by changes in the oxidation state, symmetry, and spin state of an active transition-metal center. Study of the local electronic structure is necessary to understand catalytic function so that new industrial processes can be developed and enhanced.\cite{Ostrom2015, Chabera2017} The study of biocatalysts such as enzymes is especially important to further understand disease pathways,\cite{Faller2012} develop new drugs,\cite{Liu2010} and create bio-mimetic catalysts\cite{Fukuzumi2013} that can be used in industry.
 
Core-level X-ray absorption spectroscopy (XAS), where an incident beam of X-rays is varied in energy to excite core electrons into unoccupied valence states, is a powerful tool to decipher the local electronic structure of materials with atomic specificity. The core-to-valence transition obeys energy and momentum conservation and strong selection rules that display sensitivity to the oxidation state, spin state, and local symmetry of the excited atom and its chemical surrounding.\cite{stohr1992,DeGroot2008} 

In the soft-X-ray regime (100--2000~eV), XAS has been widely adopted for the light-element K edges (1s$\rightarrow$2p) and 3d-transition-metal L edges (2p$\rightarrow$3d) due to the dipole transition into the valence orbitals of interest.\cite{stohr1992,DeGroot2005}
Although the ``textbook" XAS, i.e.\ the photoabsorption cross-section per atom, can be determined via transmission through a thin film, XAS spectra are often more conveniently collected via detection of secondary processes involving either electron or photon decay channels as an indicator of X-ray absorption. In the soft-X-ray regime, most measurements are performed via detection of electrons ejected from the sample (electron yield or EY), due to the abundance of ejected electrons and the relative ease of detection for conductive samples (a drain-current measurement via a sensitive ammeter often suffices). However, electrons have a very short mean free path in matter and are strongly impacted by electric and magnetic fields, so EY detection is surface-sensitive and requires conductive samples to avoid buildup of a surface charge that can severely distort or eliminate the XAS signal.\cite{Kelly2010} As a result of these limitations, fluorescence yield (FY) detection is a more natural tool to probe samples that are non-conductive, require in-situ setups that electrons cannot penetrate, or for which bulk sensitivity is desired. Unfortunately, the L-edge FY signal is hundreds of times weaker\cite{krause1979} than the EY signal, leading to low FY count rates. Solid-state detectors such as silicon-drift detectors (SDDs) and charge-coupled devices can be used to obtain acceptable count rates from dilute systems,\cite{George1993, Aziz2009} but are limited to energy resolution on the order of 50~eV (FWHM) at 700~eV.\cite{Struder2000,Amptek} These detectors, which have sensitivity to all X-ray energies, can obtain the total-fluorescence-yield (TFY) spectrum via collection of all X-ray emission from the sample, but for dilute compounds the TFY signal may be dominated by emission from the background matrix of the sample rather than the atomic center of interest. In this case, it is customary to collect a partial-fluorescence-yield (PFY) spectrum that consists only of photons emitted by the desired atomic center at specific emission energies. This results in an improved signal-to-background ratio, but for low-concentration samples, in particular for measurement of 3d transition-metal L edges in an oxygen-rich matrix, the poor energy resolution of a solid-state detector can lead to a background in PFY spectra that is prohibitively high. 

In resonant inelastic X-ray scattering (RIXS), the X-ray emission spectrum is recorded as a function of the exciting X-ray energy, providing access to resonantly excited X-ray emission spectra as well as low-energy loss features relative to the elastic peak.\cite{DeGroot2008,Kotani2001,Rubensson2000} Combining bright light sources with high resolution X-ray emission spectrometers approaching 10~meV resolution\cite{ghiringelli2006,dvorak2016} has allowed the dispersion of many elementary low-energy excitations to be resolved, including spin excitations and magnons,\cite{Dean2012,Schlappa2012} and even vibrational manifold mapping onto nuclear wavepackets.\cite{Pietzsch2011} 
The ability to access the orbital excitations such as d-d transitions\cite{Butorin1996} and charge transfer excitations\cite{Augustin2016} has allowed L-edge RIXS to yield more rich chemical insight into e.g.\ crystal field, charge transfer, and valency than XAS alone.\cite{VanSchooneveld2013} This has opened up more incisive characterization of frontier-orbital interactions in model catalysts, such as the real-time mapping of ligand exchange and associated spin-state dynamics in iron pentacarbonyl.\cite{Wernet2015}

For the orbital and charge-transfer excitations of outstanding interest in chemistry, a modest requirement on the X-ray emission resolution of about 1~eV is often targeted, but high throughput becomes critical. Classic Rowland-circle grating spectrometers\cite{Nordgren1989} can measure PFY-XAS and RIXS spectra with very good energy resolution, but their low throughput and requirement of a tightly focused beam can lead to long measurement times and sample damage, especially for dilute samples. Recognition of these difficulties has led to efforts to design variable-line-spacing gratings or reflection zone plates that have an order of magnitude higher throughput.\cite{Chuang2005,Qiao2017,Yin} However, to measure RIXS spectra for the most dilute samples some groups have found it necessary to build custom high-throughput instruments that efficiently target only a single element such as manganese\cite{Mitzner} or iron.\cite{Warwick2014} While this approach has broadened the reach of soft-X-ray spectroscopy, the custom-made nature of these solutions has limited their adoption.

Radiation sensitivity adds a constraint to the study of transition-metal sites, because many samples can only be exposed for a finite time before significant damage occurs; such damage often results in reduction of the active site. Sample damage has been observed to be linearly dependent on X-ray dose,\cite{VanSchooneveld2015} so the brightest, most focused beamlines will damage samples the fastest. Several groups at X-ray free-electron lasers and synchrotrons have used liquid sample jets with diffraction gratings to overcome damage when measuring dilute transition-metal solutions.\cite{Mitzner,Kunnus2016,Wernet2015} The jet ensures a continuous supply of undamaged sample which in turn allows the long measurement times needed to compensate for the low throughput of the grating. However, these experiments require a very bright synchrotron beamline or free-electron laser, a complicated liquid jet, a long measurement time, and a large sample volume, rendering them out of reach for many experimenters.
 
These challenges have prevented L-edge soft-X-ray spectroscopy from being applied to many dilute, radiation-sensitive systems. Instead, the vast majority of studies use hard X-rays ($> 4$~keV) to probe K edges, which involve the promotion of 1s core electrons.\cite{Glatzel2005} There are many experimental advantages to working with hard X-rays; K edges have significantly higher fluorescence yields than the corresponding L edges,\cite{krause1979} the emission can be efficiently monochromatized with Bragg crystals, and hard X-rays are much more penetrating than soft X-rays, which makes experiments less sensitive to the light element matrix, the sample environment, and X-ray windows. Although it is experimentally convenient to use hard X-rays, especially for dilute samples, K-edge XAS suffers from intrinsic energy broadening caused by the short lifetime of the 1s core hole ($>1$~eV for transition-metal K edges, compared to $<0.2$~eV for the corresponding L edges). In 1s2p RIXS, which involves a K-edge 1s$\rightarrow$3d absorption followed by the detection of 2p$\rightarrow$1s emission photons,\cite{Baker2017} inelastic losses corresponding to L-edge excitations can be analyzed. Here the energy broadening in the emission direction is set by the lifetime of the $2p^5 3d^{n+1}$ final state configuration, which is the same as the L-edge XAS final state.\cite{Glatzel2005} Although 1s2p RIXS has been used in place of L-edge XAS for samples which are difficult to measure with soft X-rays, it is sensitive to effects such as transition metal 3d 4p mixing as well as the tail of the intense 3p dipole transitions, which can substantially complicate the interpretation of the 1s2p RIXS spectrum, and direct measurement of L-edge XAS and RIXS spectra is preferable.

We have commissioned an array of transition-edge-sensor (TES) detectors\cite{Doriese2017} as a spectrometer for XAS and RIXS. TES detectors are energy-dispersive (ED) devices, meaning that a TES detector directly measures the energy of each incident X-ray photon. TES detectors have much better spectral resolution than solid-state detectors. Unlike wavelength-dispersive grating spectrometers, for which good energy resolution depends on a narrow entrance slit or focused beam and small acceptance angle, an array of TES detectors provides a large collection area, high quantum efficiency, and does not require a focused beam. TES arrays can therefore attain both a higher photon throughput and lower radiation dose rate than diffraction gratings.\cite{Uhlig2015} The combination of sufficient energy resolution to study most orbital and charge-transfer excitations and high throughput allows practical PFY-XAS and RIXS measurements on dilute, damage-sensitive samples. In this paper we present measurements on frozen solutions of aqueous ferricyanide (\ce{K3[Fe^{III}(CN)6}]) to demonstrate the ability of the TES array to obtain  PFY-XAS and RIXS spectra from undamaged samples as dilute as 0.5~mM. 
 
\section{Methods}
\subsection{Instrument}

We have developed and fielded an energy-dispersive soft-X-ray spectrometer for use at the Stanford Synchrotron Radiation Lightsource (SSRL) beamline 10-1. This spectrometer is based on an array of transition-edge-sensor detectors that directly measure the energy deposited by incident X-rays. Each TES detector is a cryogenically cooled, temperature-dependent resistor. When an X-ray hits a detector, the TES heats and the resistance increases sharply. The resistance change causes a current change, which can in turn be read out by sensitive cryogenic amplifiers.\cite{Doriese2016} Signals from the TES are processed with the techniques described by \citeauthor{fowler2016}\cite{fowler2016} to determine the energy of each detected X-ray.
 
The TES array at beamline 10-1 consists of 220 operational detectors, each of which has an effective area of 104~$\mu$m by 84~$\mu$m. The total active area is 1.9~mm$^2$, and the array can be positioned as close as 30~mm to the sample, and so spans a maximum solid angle of 0.002~sr. The array has been operated at output count rates up to 10,000 counts per second. The energy resolutions of the detectors in this array range from 1.5~eV to 2~eV at 700 eV.\cite{Doriese2017} As described by \citeauthor{fowler2017},\cite{fowler2017} the instrumental energy-response function of the detectors is a Gaussian with an exponentially decaying tail to low energies.  

The TES allows substantial reduction of noise in an absorption spectrum. In a FY-XAS scan, the “noise” is primarily set by counting statistics of signal and background counts. We define $N_{\rm sig}$ for a given monochromator step to be the rate of counts of fluorescent emission collected by the detector in the X-ray line(s) of interest, while $N_{\rm back}$ for that monochromator step is the detected rate of all other X-rays. The total number of signal counts is thus $N_{\rm sig}*t$ and the total background is $N_{\rm back}*t$, where $t$ is the integration time in a given monochromator step. Background X-rays are generally fluorescent emission from other atoms, but can also be X-rays from the elastically scattered beam or X-rays from other rarer events. Assuming Poisson arrival statistics and $N_{\rm sig}t + N_{\rm back}t \gg 1$, the signal-to-noise ratio, or SNR, per monochromator step is then given by:
\begin{equation} {\rm SNR} = \frac{N_{\rm sig}t}{\sqrt{N_{\rm sig}t + N_{\rm back}t}} = \frac{N_{\rm sig} \sqrt{t}}{\sqrt{N_{\rm sig} + N_{\rm back}}}.\end{equation}
Typically, $N_{\rm  sig}/N_{\rm back}$ is set by the sample and the ability of the detector to reject background emission.
In a TFY-XAS scan, in which all emitted X-rays are counted, good SNR can be obtained if $N_{\rm sig}\sqrt{t}$ is made very large, so that $N_{\rm sig}\sqrt{t} \gg \sqrt{N_{\rm sig} + N_{\rm back}}$. However, for a dilute sample, $N_{\rm back}$ may be much larger than $N_{\rm sig}$, so $\sqrt{N_{\rm sig} + N_{\rm back}} \approx \sqrt{N_{\rm back}}$. Defining $\alpha \equiv \sqrt{N_{\rm sig}}/\sqrt{N_{\rm back}}$, we see that for $\alpha \ll 1$, 
\begin{equation} {\rm SNR} \approx \alpha\sqrt{N_{\rm sig}t}. \end{equation}
If $\alpha$ is very small, it is difficult to increase either $N_{\rm sig}$ or $t$ enough to achieve high SNR, because SNR only increases as the square root of either of these quantities. In a PFY-XAS scan detected by a typical ED spectrometer such as an SDD, which has energy resolution no better than $\Delta E_{\rm FWHM} = 50$ eV in the soft-X-ray energy range, the argument is similar: the efficiency of the ED detector allows high $N_{\rm sig}$, but $N_{\rm back}$ may still be large, especially in dilute samples.
A PFY-XAS scan acquired by a grating spectrometer, by contrast, can reduce $N_{\rm back}$ to nearly zero because background emission is separated in energy from signal emission. This is valuable to increase SNR, especially in dilute samples, as in this case \begin{equation} {\rm SNR}
\approx \frac{N_{\rm sig} t}{ \sqrt{N_{\rm sig}t }} = \sqrt{N_{\rm sig} t}.\end{equation} The drawback of grating spectrometers is their low collection efficiency, which results in a very small $N_{\rm sig}$ and increases the time needed to achieve a given SNR. This can push achievement of acceptable SNR out of reach for many dilute samples of interest. Our TES spectrometer, by contrast, combines the large collecting efficiency of an ED spectrometer with good energy resolution so that $N_{\rm back} \approx 0$, but $N_{\rm sig}$ is still large enough that ${\rm SNR} \approx \sqrt{N_{\rm sig} t} > 1$ for reasonable measurement times $t$.

\subsection{Experiment}

\ce{K3[Fe^{III}(CN)_6]} was purchased from Sigma Aldrich and used without further purification. Solutions with concentrations from 0.5~mM to 500~mM were prepared in deionized water. These liquid samples were then deposited onto a specially designed aluminum bar with shallow sample pockets and rapidly frozen in liquid nitrogen. This sample holder was then immediately mounted to a cryogenically cooled (LN$_2$) copper receiver, pumped to high vacuum pressures ($< 10^{-7}$ Torr), and kept at 80~K for the duration of the measurements. We did not use any window to cover the face of the frozen liquid samples. Measurements were performed on frozen aqueous solutions of 0.5~mM, 5~mM, 50~mM, and 500~mM concentration. 
 
X-ray spectra were recorded at the soft-X-ray wiggler beamline 10-1 of the Stanford Synchrotron Radiation Lightsource. The X-rays were monochromatized via a spherical grating monochromator with 1000 lines/mm and 20~$\mu$m entrance and exit slits, which provided $5\times10^{10}$ ph/s with a resolution of approximately 0.15~eV (FWHM) in a 1~mm$\times$1~mm spot on the sample. Beam damage was carefully controlled by establishment of the maximum dose before non-negligible reduction was observed in the spectra, which for our flux density meant a maximum exposure of 7 minutes per spot for \ce{K3[Fe^{III}(CN)6]}. At our observed beam flux this corresponds to a 2 MGy skin dose, which is below previously observed exposure limits for \ce{K3[Fe^{III}(CN)6]}.\cite{VanSchooneveld2015} The X-rays impinged onto the surface of the frozen solution at 45$^{\circ}$ degree incidence and emission was recorded in the horizontal plane (along the E-vector of the incoming, linearly polarized X-rays). We operated the TES detector array at a sample-to-detector distance of about 50~mm, corresponding to a solid angle of about 0.0008~sr. The beam energy was swept from 690--700~eV in 0.2~eV steps, 700--735~eV in 0.1~eV steps, 735--740~eV in 0.5~eV steps, and 740--830~eV in 2~eV steps. 

\section{Results}

\begin{figure*}
\includegraphics{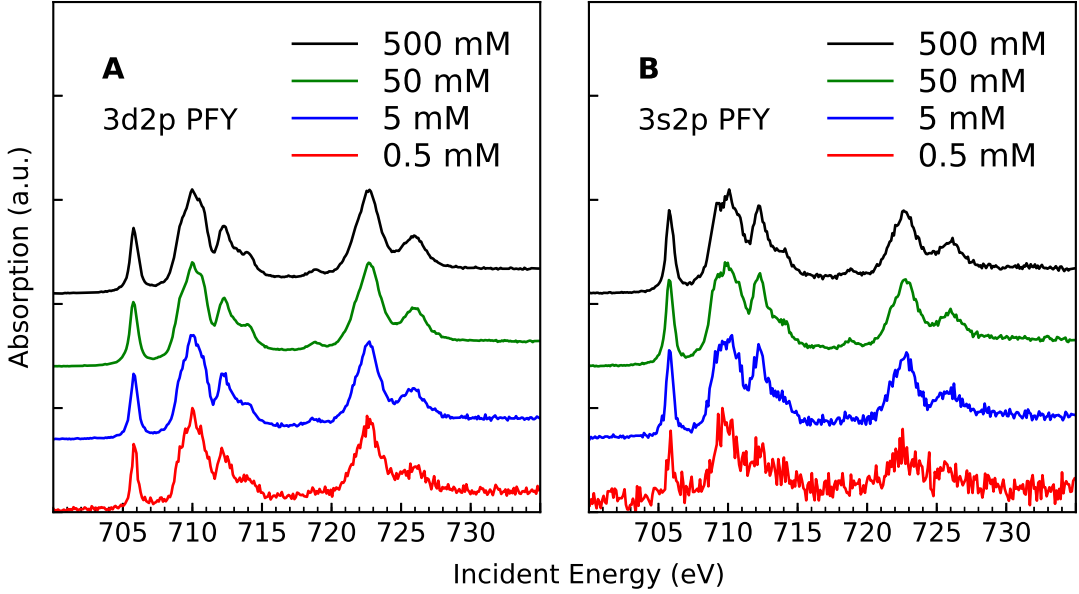}
\caption{\small Comparison of (A) Fe 3d2p PFY-XAS and (B) Fe 3s2p PFY-XAS measurements for all concentrations of aqueous \ce{K3[Fe(CN)6]} . Spectra have been normalized and vertical offsets applied for clarity. Spectra for 500 mM, 50 mM, and 5 mM concentrations were acquired in 5 hours each, while the 0.5 mM spectrum took 15 hours of measurement. The 3d2p signal has six times greater intensity than the 3s2p signal.
}
\label{fig:conc_series}
\end{figure*}

\begin{figure*}[htb!]
\includegraphics{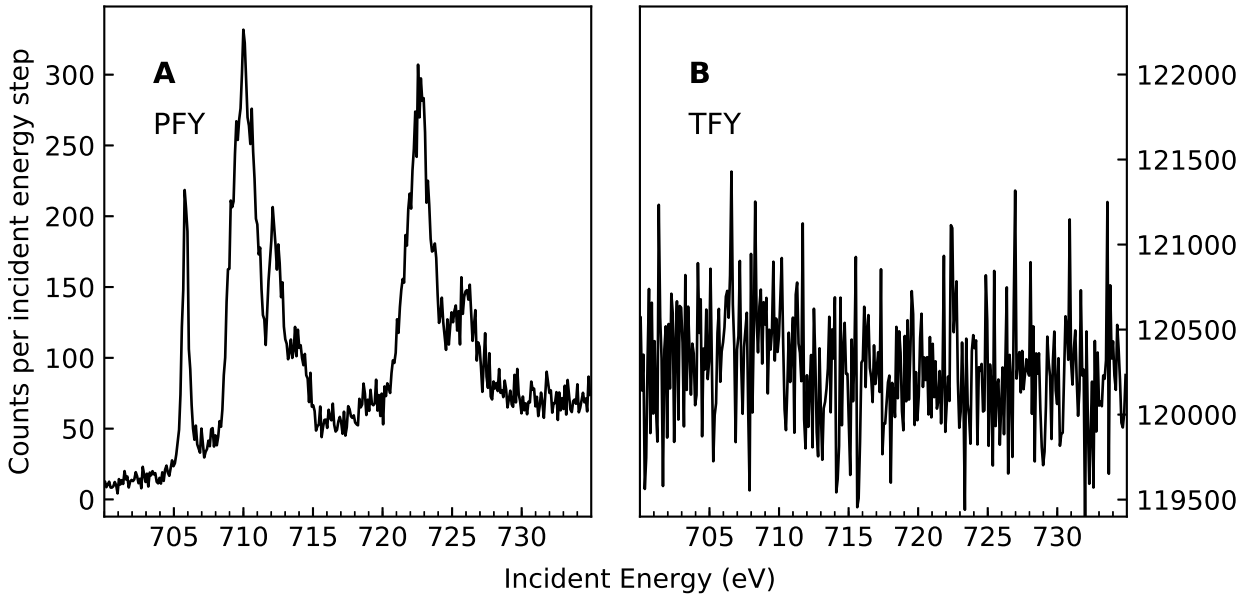}
\caption{\small XAS spectra from the 0.5 mM sample showing (A) The PFY signal with an emission energy window of 680--740 eV, and (B) the TFY signal (all emission energies). Both spectra were produced from the same data by summing the appropriate emission window.
}
\label{fig:pfy_tfy}
\end{figure*}

Figure \ref{fig:conc_series} shows both 3d2p and 3s2p PFY-XAS spectra for all sample concentrations. The 3d2p PFY-XAS spectrum contains only emission from the $2p^5 3d^{n+1}\rightarrow 2p^6 3d^n$ (3d2p) transition, following the $2p^63d^n \rightarrow 2p^5 3d^{n+1}$ L-edge absorption. 3s2p PFY-XAS contains the $2p^5 3s^2 3d^{n+1}\rightarrow 2p^6 3s^1 3d^{n+1}$ (3s2p) emission. To create the 3d2p PFY-XAS spectrum we integrated all emission from 680--840~eV, and for the 3s2p spectrum we integrated emission from 590--640~eV.\footnote{The window of 680--840 was chosen to be larger than the full range of monochromator energies so that all bins would have similar contributions from the elastically scattered beam. A narrower emission window would lead to the elastic signal entering and exiting the PFY region, which could create a spurious signal.} Our 3d2p PFY-XAS spectra have acceptable SNR even at the lowest concentration of 0.5~mM. The measured 3s2p fluorescence is a factor of six weaker than the 3d2p fluorescence, but useful 3s2p PFY-XAS is still achieved at 5~mM.\footnote{A SNR calculation using Eq.1 shows that at our present background level, a 0.5~mM spectrum requires 12 times as long a measurement as a 5~mM sample to obtain the same signal-to-noise ratio. This shows that the 0.5~mM spectrum could be brought to the noise level of the 5~mM spectrum through 60 hours of measurement--a long but feasible measurement.} Figure \ref{fig:pfy_tfy} shows two unnormalized XAS spectra from the 0.5~mM data to illustrate the levels of background counts in the PFY and equivalent TFY measurements. Before we use our detector's energy resolution to reject emission from background matrix elements, the background level ($N_{\rm back}t$) is about 120,000 counts per incident energy bin, the SNR is low as predicted by Eq. 2, and the iron L-edge spectral features are entirely obscured. After we window to select just the 3d2p emission, $N_{\rm back}$ is reduced by a factor of 10,000 and the SNR increases by a factor of 100. Figure \ref{fig:pfy_tfy}B demonstrates the challenge faced by TFY detectors; because SNR (Eq.~2) improves as the square root of the total counts, a detector collecting a TFY signal would have to collect 10,000 times as many counts as our TES detector in order to achieve the same signal-to-noise ratio. A typical grating spectrometer, on the other hand, would be expected to have the same signal-to-background level displayed in Figure \ref{fig:pfy_tfy}A, but would have to spend 100 times as long as our TES to collect that signal, due to the 100-fold lower throughput of most grating systems.\cite{Uhlig2015}

\begin{figure*}[htb!]
\includegraphics{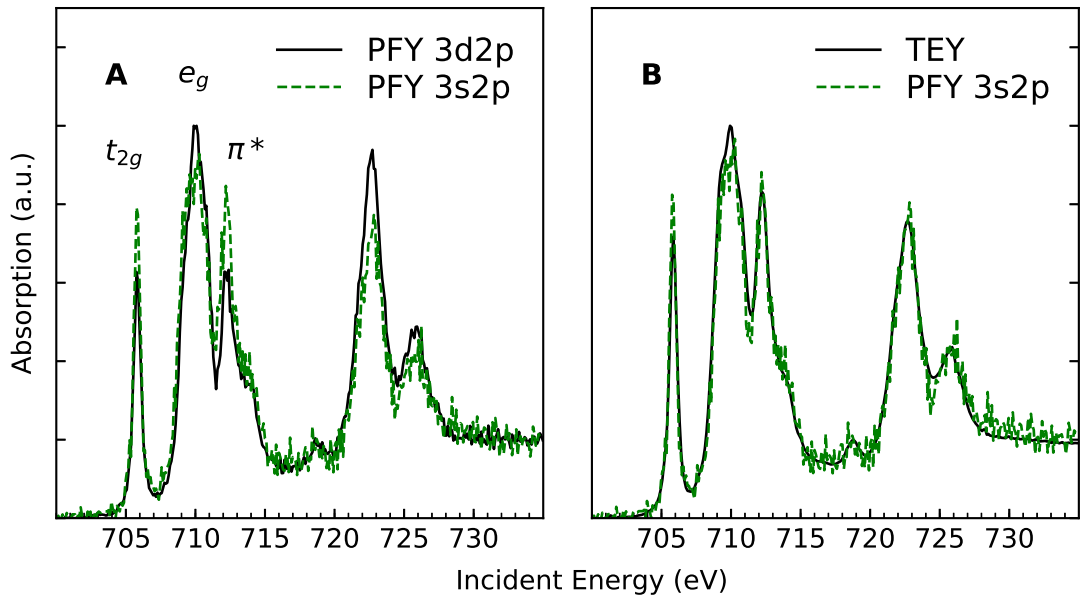}
\caption{\small Comparisons between (A) the Fe 3s2p and 3d2p PFY-XAS spectra from the 5 mM sample and (B) the 3s2p PFY-XAS spectrum from the 5 mM sample and a TEY-XAS spectrum from powdered \ce{K3[Fe(CN)6]} detected via a channeltron, also at SSRL BL10-1. Spectra were normalized to have equal areas. No extra background subtraction was done beyond the PFY windowing. The peaks of the L$_3$ edge are labeled with the transition assignments given by \citeauthor{Hocking2006}\cite{Hocking2006}}
\label{fig:pfy_tey}
\end{figure*}

Figure~\ref{fig:pfy_tey}A shows the Fe L$_3$-L$_2$ region of the 3d2p and 3s2p PFY-XAS spectra for the 500~mM \ce{K3[Fe^{III}(CN)6]} sample. Our 3s2p PFY-XAS spectrum is the first reported of this compound. The 3s2p PFY-XAS spectrum displays expected differences compared to the 3d2p: the L$_2$ edge intensity is enhanced in the 3d2p channel and the intensities of the $t_{\rm 2g}$ and  $\pi^*$ peaks are reduced. The differences between the 3d2p, 3s2p, and electron-yield spectra have been discussed in the literature. \cite{DeGroot1995,Golnak2016,miedema2014,kurian2012} When saturation effects are avoided, the 3s2p spectrum can be expected to reproduce the true X-ray absorption cross-section, which will be identical to the electron-yield spectrum. Figure~\ref{fig:pfy_tey}B shows that our 3s2p PFY spectrum is indeed a close match to a conventional TEY spectrum that we obtained on a powdered sample.

\begin{figure*}
\includegraphics{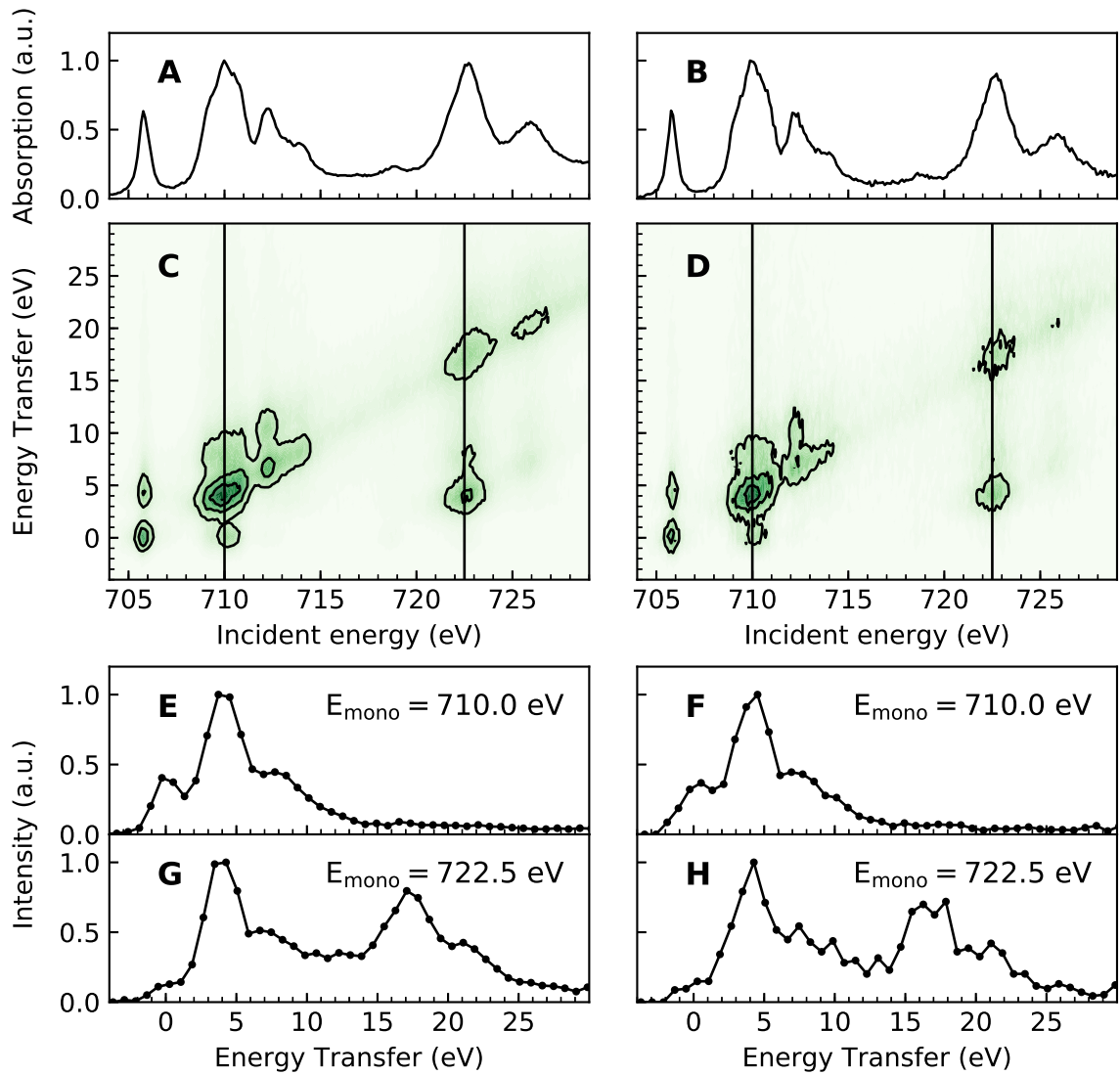}
\caption{The iron L$_3$-L$_2$ region of the RIXS plane for 500 mM (left) and 5 mM (right) samples, with emission cuts and PFY-XAS. Each RIXS plane was produced in 5 hours of measurement time. The bin size on the energy transfer axis is 0.8~eV. (A) and (B) show the PFY-XAS for each sample, produced by the direct sum of counts in (C) and (D), respectively for each excitation energy. (C) and (D) show the RIXS plane, with energy transfer (incident energy $-$ emission energy) on the vertical axis. (E), (F), (G), and (H) are vertical cuts through the RIXS plane at incident energies of 710 eV and 722.5 eV.}
\label{fig:rixs_zoom}
\end{figure*}

In figure~\ref{fig:rixs_zoom} we show a portion the RIXS planes collected from the 500~mM and 5~mM samples. These RIXS planes reproduce all of the major features of the RIXS plane published by \citeauthor{Kunnus2016}, \cite{Kunnus2016} which was produced from a liquid-jet setup with a \ce{K3[Fe^{III}(CN)6]} concentration of 500~mM. Thus we demonstrate that TES-RIXS of archetypal model compounds can replicate RIXS taken by conventional soft x-ray grating spectrometers, while the TES has the ability to probe much lower concentrations than have previously been attainable.

\section{Discussion}
Measurement of dilute samples is part of a long-standing effort by the spectroscopy community to access the electronic structure of transition-metal active sites.\cite{Aziz2009,Mitzner} As part of this effort, aqueous ferricyanide has been a prototype system for X-ray spectroscopy, especially in the study of metal-ligand charge transfer and differential orbital covalency.\cite{Hocking2006} Compounds such as ferricyanide then serve as references to infer the local electronic structure of complex metalloproteins. Despite the progress made with model compounds, the overarching goal remains to obtain spectra from actual metalloproteins and to compare these spectra to those of the model compounds. Whereas model compounds are easy to work with and can be prepared at Fe concentrations exceeding 500~mM, many important metalloproteins can only be prepared in the 0.5~mM to 5~mM concentration range, which has been a significant impediment to their study in the soft-X-ray regime. Despite the relative ease of measuring \ce{K3[Fe^{III}(CN)6]}, a full soft-X-ray RIXS plane of 500~mM ferricyanide was first measured with a grating spectrometer\cite{Kunnus2016} only in \citeyear{Kunnus2016}.  Other groups have made TFY measurements of 5~mM of iron, but have not demonstrated the ability to obtain either a PFY spectra or a RIXS plane at those low concentrations.\cite{Aziz2009} In this context, our measurement of a PFY spectrum from a 0.5~mM sample with a TES spectrometer is a substantial step toward making measurements of dilute samples a routine capability of soft-X-ray beamlines. These measurements are made possible by the unique combination of high throughput and 1.5~eV energy resolution provided by the TES spectrometer.
 
\ce{K3[Fe^{III}(CN)6]} can be damaged by prolonged exposure to X-rays, so we restricted the radiation dose delivered to the sample by using a large beam spot (1~mm$^2$) and a short scan time. A typical grating spectrometer requires a tightly focused, 10~$\mu$m$\times$100~$\mu$m spot, which yields a radiation dose rate one thousand times higher than that of a 1~mm$\times$1~mm beam. A high dose rate makes it extraordinarily difficult to complete a full monochromator energy scan on a solid sample before damage occurs, which is why grating-based measurements in the soft-X-ray regime often require a liquid jet. A liquid jet mitigates radiation damage, but restricts the application of soft-X-ray L-edge measurements to samples in liquid solution that can be prepared in large volumes. Because the TES is energy-dispersive, not wavelength-dispersive, the beam spot size does not affect its energy resolution. For dilute solid samples that are even more sensitive to X-ray damage than \ce{K3[Fe^{III}(CN)6]}, we could defocus the X-ray beam to a 10~mm$^2$ spot, which would yield even lower photon-induced damage without loss of data quality or increase in acquisition time. This is an enabling capability for static radiation-sensitive samples.

Our TES spectrometer measures the 3d2p and 3s2p emission simultaneously. As presented in figure~\ref{fig:pfy_tey}, the 3s2p spectrum is a close match to the TEY spectrum, whereas the 3d2p spectrum contains distortions due to ``state-dependent decay.''\cite{DeGroot1995,kurian2012,miedema2014} The 3s2p PFY-XAS is, barring saturation effects, a direct representation of the XAS cross-section, much like the TEY-XAS mode is for appropriately concentrated and conductive samples. Additionally, the 3s2p PFY-XAS measurement does not suffer from self-absorption effects like the 3d2p can, because the 3s2p emission is below the iron edge and cannot be resonantly re-absorbed. Despite the advantages of 3s2p measurements, published examples of 3s2p spectra are rare because the lower intensity of the 3s2p emission leads to time-consuming data collection with a traditional grating spectrometer. Because the TES spectrometer collects all X-ray energies simultaneously, TES data sets automatically contain the 3s2p emission spectrum, which will be an invaluable proxy for the ``textbook'' XAS cross-section for low-concentration samples from which TEY-XAS cannot ordinarily be measured.

\section{Conclusion}

We have collected L-edge XAS and RIXS spectra on frozen samples of aqueous \ce{K3[Fe^{III}(CN)6]} at concentrations down to 0.5~mM. Our spectra accurately reproduce existing measurements\cite{Kunnus2016} of high-concentration aqueous \ce{K3[Fe^{III}(CN)6]}. The TES array is able to collect spectra at these low concentrations due to the unique combination of high efficiency and sufficient energy resolution. The TES has several distinct advantages over traditional spectrometers that have been used to measure L-edge spectra of dilute biological samples. The TES spectrometer is able to probe lower sample concentrations, collect a wide range of emission energies, and measure solid, radiation-sensitive samples. Unlike a grating spectrometer, the TES is able to measure with a large x-ray beam spot to prevent sample damage without sacrificing energy resolution or acquisition time. These advantages will allow us to measure important metalloproteins that are susceptible to damage and cannot be concentrated more than a few mM. The application of RIXS and PFY-XAS to dilute, radiation-sensitive, static samples constitutes one of the most important and transformative opportunities enabled by TES technology.

\begin{acknowledgments}

This work is supported by the Department of Energy, Laboratory Directed Research and Development funding, under contract DE-AC02-76SF00515. Use of the Stanford Synchrotron Radiation Lightsource, SLAC National Accelerator Laboratory, is supported by the U.S. Department of Energy, Office of Science, Office of Basic Energy Sciences under contract No. DE-AC02-76SF00515. M.L.B. acknowledges the support of the Human Frontier Science Program. NIST authors acknowledge support from the NIST Innovations in Measurement Science Program and the DOE Office of Basic Energy Sciences.
This research was supported by the National Institute of General Medical Sciences of the National Institutes of Health under award number R01GM040392 (E.I.S.) K.M.M. acknowledges the support of a National Research Council Postdoctoral Fellowship.
\end{acknowledgments}

\bibliography{fecn6_paper}

\end{document}